\documentclass[pra,twocolumn]{revtex4-1}
\usepackage{hyperref} 

\usepackage{graphicx} 
\usepackage{subfig}
\usepackage{color}
\usepackage{filecontents}
\usepackage{soul}
\usepackage{mathrsfs}
\usepackage{tikz,bm} 
\usepackage{circuitikz}
\usepackage{tikz}
\usepackage{physics}
\usepackage{mathtools}
\usepackage[T1]{fontenc}
\usepackage[outline]{contour}
\usepackage{relsize}

\usetikzlibrary{shapes.geometric}
\usetikzlibrary{decorations.pathmorphing}
\usetikzlibrary{shapes.arrows, fadings}
\usetikzlibrary{arrows,snakes,shapes}
\usetikzlibrary{positioning}
\definecolor{mycolor}{rgb}{1,0.2,0.3}
\definecolor{brightgreen}{rgb}{1.0, 1.0, 1.0}
\definecolor{britishracinggreen}{rgb}{0.0, 0.26, 0.15}
\definecolor{cadmiumgreen}{rgb}{0.0, 0.42, 0.24}
\definecolor{ceruleanblue}{rgb}{0.16, 0.32, 0.75}
\definecolor{darkelectricblue}{rgb}{0.33, 0.41, 0.47}
\definecolor{darkpowderblue}{rgb}{0.0, 0.2, 0.6}
\definecolor{dt}{rgb}{1.0, 0.66, 0.07} 
\definecolor{emerald}{rgb}{0.31, 0.78, 0.47}
\definecolor{palatinatepurple}{rgb}{0.41, 0.16, 0.38}
\definecolor{pastelviolet}{rgb}{0.8, 0.6, 0.79}
\definecolor{br}{rgb}{0.5, 0.05, 0.01}
\definecolor{chosen_color}{RGB}{3, 207, 252}

\pgfdeclareverticalshading{pgf@lib@fade@north}{100bp}
{color(0bp)=(pgftransparent!0);
 color(30bp)=(pgftransparent!0);
 color(50bp)=(pgftransparent!80); 
 color(65bp)=(pgftransparent!100);
 color(100bp)=(pgftransparent!100)}%
\pgfdeclarefading{exponentialtop}{%
\pgfuseshading{pgf@lib@fade@north}%
}

\pgfdeclareverticalshading{pgf@lib@fade@north}{100bp}
{color(0bp)=(pgftransparent!100);
 color(35bp)=(pgftransparent!100);
 color(50bp)=(pgftransparent!80); 
 color(70bp)=(pgftransparent!00);
 color(100bp)=(pgftransparent!00)}%
\pgfdeclarefading{exponentialbottom}{%
  \pgfuseshading{pgf@lib@fade@north}%
}

\newcommand{\be}{\begin{equation}}
\newcommand{\ee}{\end{equation}}
\newcommand{\bea}{\begin{eqnarray}}
\newcommand{\eea}{\end{eqnarray}}
\newcommand*{\myeqref}[2][Eq.~]{%
\hyperref[{#2}]{#1(\ref*{#2})}%
}
\def\equationautorefname#1#2\null{%
Eq.#1(#2\null)%
}

\usepackage{dcolumn}
\usepackage{bm}
\usepackage{amssymb,amsmath}
\usepackage{color}
\usepackage{float}
\usepackage{tikz}
\usetikzlibrary{arrows}

\definecolor{DarkGreen}{rgb}{0,0.6,0.2}

\usepackage{mathrsfs}

\begin{document}
\title{Chirality-free photon routing via giant atoms in waveguide QED ladders}
\author{Glenn Ochsner$^{1}$} 
\author{Vincent Schena$^{1}$} 
\author{Arushi Deb$^{1}$} 
\author{Imran M. Mirza$^{1,2}$}
\email{mirzaim@miamioh.edu}
\affiliation{$^{1}$Macklin Quantum Information Sciences, Department of Physics, Miami University, Oxford, OH 45056, USA\\
$^{2}$Department of Computer Science \& Software Engineering, Miami University, Oxford, OH 45056, USA}
\date{\today}

\begin{abstract}
In this paper, we present an in-depth examination of single-photon routing in a multi-emitter waveguide quantum electrodynamics ladder with up to five giant atoms simultaneously coupled across two linear waveguides. Using a real-space approach, we analyze a non-chiral routing architecture and evaluate the impact of scaling the number of giant atoms in three topologically distinct configurations: fully separated, simply braided, and fully nested. Using numerical results, we show that increasing the number of giant atoms, in separate or braided configurations, greatly expands the operational parameter space for near-deterministic ($\sim$100\%) photon routing into the target waveguide, achieving directionality through quantum interference alone. In contrast, the nested architecture is limited to a maximum routing efficiency of 25\% due to rigid geometric symmetries, although this limitation could be mitigated by employing less symmetric coupling distances. Finally, we assess the system under direct interatomic interactions and environmental dissipation. Our results show that, while spontaneous emission reduces transport probabilities, the multi-atom-separated configuration maintains high routing efficiencies across broad operational windows of incoming photon frequency without requiring chiral coupling (or the light’s spin-momentum locking mechanism).
\end{abstract}


\maketitle

\section{\label{sec:I} Introduction}  
Over the past few decades, superconducting quantum circuits have emerged as one of the leading hardware platforms for scalable quantum computing and quantum information processing \cite{wendin2017quantum}. In these architectures, generating, manipulating, and routing single photons with high fidelity is a foundational requirement for building distributed quantum networks and on-chip quantum optics \cite{kimble2008quantum}. Waveguide quantum electrodynamics (wQED) offers a powerful framework for these tasks by studying the interactions between quantum emitters and photons propagating through one-dimensional transmission lines \cite{sheremet2023waveguide, roy2017colloquium, blais2021circuit, you2011atomic}. Unlike traditional cavity QED, where photons are trapped inside high-finesse optical cavities \cite{walther2006cavity}, wQED configurations enable long-range, directional, and waveguide-mediated interactions among multiple localized emitters. This makes them uniquely suited for scalable quantum state transport \cite{roy2017colloquium, zheng2013waveguide}. 

Traditionally, quantum emitters in wQED have been treated as ``small atoms'' under the dipole approximation \cite{gerry2023introductory}, assuming their effective physical size to be negligible compared to the wavelength of the interacting photon mode. In this regime, the atom couples to the waveguide at a single point. A major paradigm shift occurred with the realization of ``giant atoms'' in superconducting circuit platforms \cite{kannan2020waveguide}. Giant atoms couple to the linear waveguide at multiple, spatially separated points. When the distance between these points is comparable to the wavelength of the guided mode, non-local interactions introduce substantial phase delays and quantum interference effects \cite{frisk2020quantum}. This self-interference fundamentally modifies the scattering, emission, and bound-state properties of the system, enabling a number of unique features such as tunable nonreciprocal transmission \cite{chen2022nonreciprocal}, subradiance \cite{qiu2023collective}, and engineered photon-photon correlations \cite{gu2024tunable, gu2023correlated}.

We and others have reported the possibility of deterministic single-photon routing in small atom waveguide QED configurations \cite{gonzalez2016nonreciprocal, poudyal2020collective}. The essential idea in those studies was to leverage the spin-momentum locking mechanism, in which small or point-like quantum emitters couple to engineered nanophotonic paths in which the local polarization state (spin) of light is locked to its propagation direction (momentum) \cite{lodahl2017chiral, suarez2025chiral}. While these chiral interfaces break time-reversal symmetry to guide photons unidirectionally with high precision, they often require complex material engineering, precise magnetic-field tuning, or sub-wavelength positioning. Their routing efficiency can also be limited by environmental dissipation and placement disorders, which must be countered with collective atomic effects (see, for example, Ref.~\cite{amgain2024photon}). This shifts the focus back to geometric optimization in bidirectional, multi-point systems where interference alone can orchestrate routing behavior.

Despite rapid theoretical and experimental progress in giant-atom architectures, the literature remains focused on minimal setups. Most studies analyze either a single giant atom or pairs of giant atoms coupled to one or two waveguides \cite{wang2025targeted, wang2021giant, zhang2022controllable, sun2025frequency, chai2025single}. While these models demonstrate basic routing, the scalability, geometric optimization, and structural limits of multi-atom arrays within multi-waveguide routers remain largely unexplored. Understanding how the number of emitters scales with their relative topology is essential for building robust, high-efficiency quantum routing networks. In this work, we address this gap by employing the real-space approach of quantum optics \cite{shen2005coherent, berndsen2024few, mirza2017chirality} to analyze single-photon routing across different network topologies. Specifically, we investigate a single-photon router comprising up to five giant atoms coupled across two independent linear waveguides. We systematically evaluate three distinct spatial configurations of the coupling nodes: {\it fully separated}, {\it simply braided}, and {\it fully nested topologies}.

Our findings reveal that increasing the number of giant atoms in separate or braided configurations greatly enhances routing efficiency, achieving near-deterministic ($\sim$100\%) photon transmission into the target waveguide. The nested configuration, however, suffers from intrinsic geometric symmetries that limit its maximum routing efficiency to 25\%, making it ineffective for scaling. We also test practical limits by incorporating direct interatomic coupling schemes and accounting for environmental losses due to spontaneous atomic emission. Our results show that although environmental dissipation degrades routing performance, the separated multiple-giant-atom configuration retains robust operational windows. This highlights its potential as a flexible yet resilient setup for quantum routing devices that does not require explicit chiral coupling mechanisms.

The remainder of this paper is structured as follows. In Section~\ref{sec:II}, we present the analytical framework and routing limits for a single giant atom setup. Next, in Section~\ref{sec:III}, we evaluate the two-giant-atom system across separate, braided, and nested configurations. Afterward, in Section~\ref{sec:IV}, we expand this analysis to multi-atom networks of three, four, and five giant atoms, highlighting the scaling advantages and the impact of interatomic interactions. In Section~\ref{sec:V}, we introduce environmental loss and examine the structural resilience of each configuration against spontaneous emission. Finally, in Section~\ref{sec:VI}, we close with a brief summary of this paper and concluding remarks.

\section{\label{sec:II} Routing with a single giant atom}
\begin{figure}
\includegraphics[width=2.65in, height=1.95in]{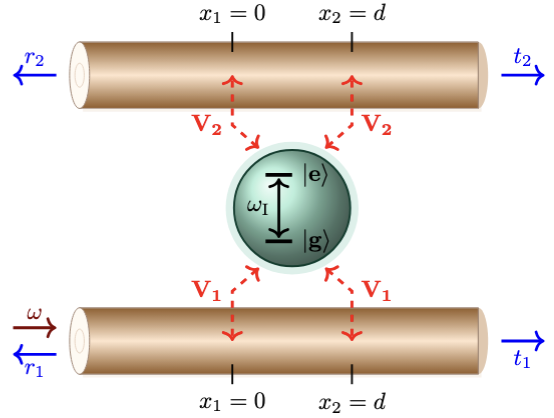} 
\captionsetup{
format=plain,
margin=1em,
justification=raggedright,
singlelinecheck=false
}
\caption{(Color online) Single photon router using a single giant atom, coupled to each waveguide at $x_{1} = 0$ and $x_{2} = d$. The photon enters the setup from the bottom left.}\label{1Atom}
\end{figure}
We begin our study of a ladder-based single-photon router with the simplest case: a single two-level giant atom. Figure~\ref{1Atom} shows the setup. The giant atom has states $|g\rangle$ and $|e\rangle$ and couples to two bidirectional waveguides with linear dispersions labeled 1 and 2. Although a giant atom can couple to each waveguide at any number of points, we consider configurations where it couples at exactly two spatially separated points for each waveguide: $x_{1} = 0$ and $x_{2} = d$. We assume the spacing between these points is constant for both waveguides, even when coupling to more than one giant atom (see next Sections for further details). Assuming equal coupling strength to each waveguide ($V=V_1=V_2$), the Hamiltonian (with $\hbar = 1$) is given by
\begin{align} \label{H-1}
&\hat{\mathcal{H}} = \omega_\text{I} \hat{\sigma}_{\text{I}}^\dagger \hat{\sigma}_{\text{I}} -i v_g \sum^2_{q=1} \int^{+\infty}_{-\infty} dx \big[ \hat{c}^\dagger_{Rq}(x) \partial_x\hat{c}_{Rq}(x) - \nonumber\\ & \hat{c}_{Lq}^\dagger(x) \partial_x \hat{c}_{Lq}(x)\big] + \sum_{p \in \{L,R\}} \sum^2_{q = 1} V \big[ \left(\hat{c}_{pq}^\dagger(0) + \hat{c}_{pq}^\dagger(d)\right)\hat{\sigma}_{\text{I}}\nonumber\\
&+ \hat{\sigma}^\dagger_{\text{I}}\left(\hat{c}_{pq}(0) + \hat{c}_{pq}(d)\right) \big]. 
\end{align} 
Here, $\omega_\text{I}$ is the transition frequency of the giant atom, $\hat{\sigma}_{\text{I}}^\dagger$ ($\hat{\sigma}_{\text{I}}$) is the raising (lowering) operator, and $v_g$ is the photon group velocity, assumed identical in both waveguides. The operator $\hat{c}_{pq}^\dagger$ ($\hat{c}_{pq}$) for $p \in \lbrace L,R\rbrace$ and $q = 1,2$ creates (annihilates) a right- or left-traveling photon in waveguide $q$.

Since there is only one photon in the system, the quantum state restricted to the single excitation sector of the Hilbert space can be expressed as
\begin{align}\label{psi-1}
|\psi\rangle = & \left\lbrace u_\text{I}\hat{\sigma}_{\text{I}}^\dagger + \sum_{p \in \{L,R\}} \sum^2_{q=1} \int^{+\infty}_{-\infty} \left[ \varphi_{pq}(x) \hat{c}_{pq}^\dagger (x) \right] \right\rbrace |\varnothing\rangle,
\end{align}
where $u_\text{I}$ is the probability amplitude when the giant atom is excited, and there are no photons in the waveguides. $\varphi_{Lq}(x)$(or $ \varphi_{Rq}(x))$ denotes the probability amplitude when the single photon is in the $q$th waveguide traveling in the left (or right) direction, and the giant atom is in its ground state. $|\varnothing\rangle$ represents the global ground state of the system when there are no photons in the waveguide and the giant atom is unexcited.

\begin{figure*} 
\begin{centering}
\begin{tikzpicture}
    \node (R) [rotate = 90] {Reflection};
    \node (R1) [right of = R, xshift = 0.11\textwidth, yshift = -0.02\textwidth] {\includegraphics[width = 0.3\textwidth, trim={30 0 0 0}, clip]{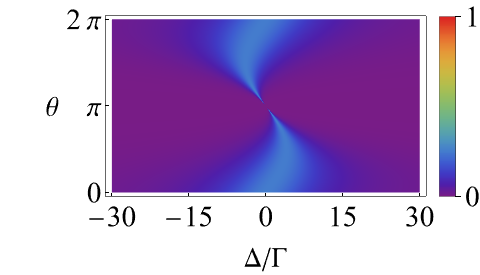}};
    \node (R2) [right of = R1, xshift = 0.25\textwidth] {\includegraphics[width=0.3\textwidth, trim = {30 0 0 0}, clip]{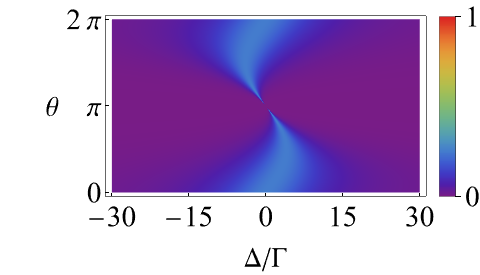}};
    \node (W1) [above of = R1, yshift = 0.04\textwidth, xshift = 0.004\textwidth] {Waveguide 1};
    \node (W2) [above of = R2, yshift = 0.04\textwidth, xshift = 0.004\textwidth] {Waveguide 2};
     \node (T) [below of = R, yshift = -0.13\textwidth, rotate = 90] {Transmission};
    \node (T1) [below of = R1, yshift = -0.13\textwidth]{\includegraphics[width=0.3\textwidth, trim = {30 0 0 0}, clip]{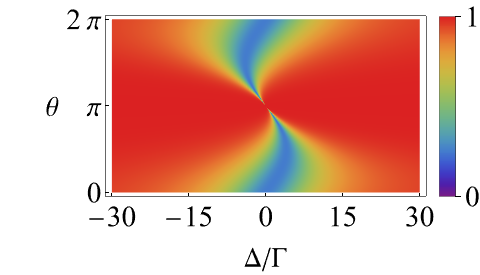}};
    \node (T2) [below of = R2, yshift = -0.13\textwidth]{\includegraphics[width=0.3\textwidth, trim = {30 0 0 0}, clip]{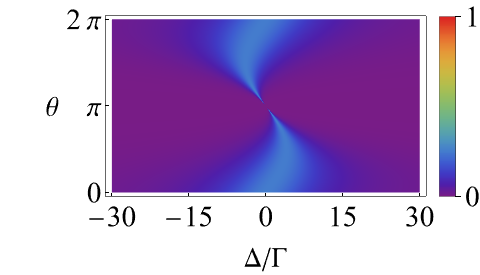}};
    \end{tikzpicture}
\captionsetup{
format=plain,
margin=1em,
justification=raggedright,
singlelinecheck=false
}
\caption{(Color online) Photon transmission and reflection probabilities in waveguides 1 and 2 for a single giant atom. Detuning $\Delta$ is varied from $-30$ to $+30$ in units of the atom-waveguide coupling rate $\Gamma$. The phase $\theta$, which incorporates the separation $d$ between coupling points, is varied from $0$ to $2\pi$.}
\label{singleAtomPlots}
\end{centering}
\end{figure*}
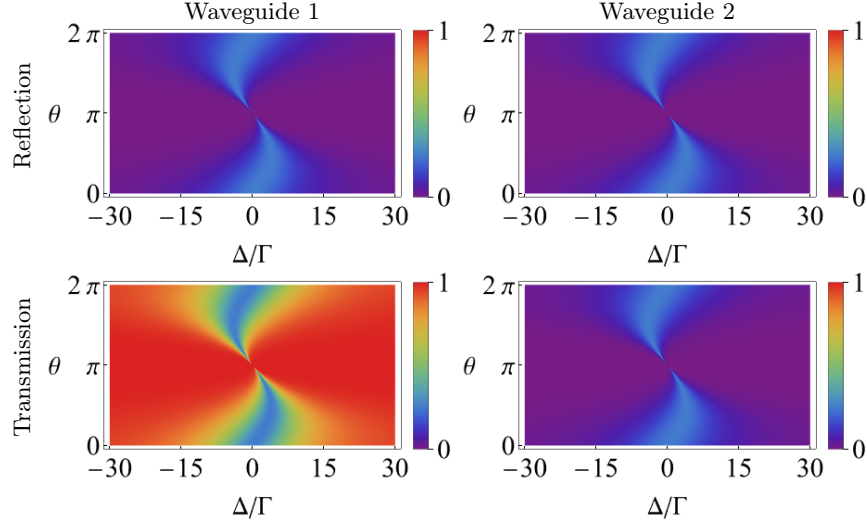

Next, we insert Eq.~(\ref{H-1}) and Eq.~(\ref{psi-1}) into the time-independent Schr\"odinger equation $\hat{\mathcal{H}}\ket{\psi}=\omega\ket{\psi}$, where $\omega$ is the frequency of the photon launched from the bottom left port of the routing setup (see Fig.~\ref{1Atom}). As a result, we obtain the following set of equations obeyed by the probability amplitudes
\begin{subequations}
  \begin{align}
    & u_{\text{I}}\Delta - \sum^{2}_{q=1} V_q \left[ \varphi_{Rq}(0) + \varphi_{Rq}(d) + \varphi_{Lq}(0) + \varphi_{Lq}(d) \right] =0, \\
    & -i v_g \frac{d\varphi_{Lq}(x)}{dx} = \omega \varphi_{Lq}(x) - V_q u_{\text{I}} \left[ \delta(x) + \delta(x-d) \right],\\
    & i v_g \frac{d\varphi_{Rq}(x)}{dx} = \omega \varphi_{Rq}(x) - V_q u_{\text{I}} \left[ \delta(x) + \delta(x-d) \right],
\end{align}
\end{subequations}
where $\Delta\equiv \omega - \omega_\text{I}$. Note that given that $q=1,2$, the last two subequations correspond to four distinct equations. Following the Bethe Ansatz \cite{shen2007strongly} and regularization condition at the coupling points $x=0$ and $x=d$, we assume the waveguide probability amplitudes are of the form
\begin{subequations}
\begin{align}
    & \varphi_{L2}(x) = \begin{cases} r_2 e^{-ikx} & \text{if } x < 0 \\
        \frac{r_2 + \alpha_2}{2} & \text{if } x = 0 \\
        \alpha_2 e^{-ikx} & \text{if } 0 < x < d \\
        \frac{\alpha_2}{2} e^{-ikd} & \text{if } x = d \\
        0 & \text{if } x > d \end{cases} \\
        & 
    \varphi_{R2}(x) = \begin{cases} 0 & \text{if } x < 0 \\
        \frac{\beta_2}{2} & \text{if } x = 0 \\
        \beta_2 e^{ikx} & \text{if } 0 < x < d \\
        \frac{\beta_2 + t_2}{2} e^{ikd} & \text{if } x = d \\
        t_2 e^{ikx} & \text{if } x > d \end{cases} \\
    & \varphi_{L1}(x)  = \begin{cases} r_1e^{-ikx} & \text{if } x < 0 \\
        \frac{r_1 + \alpha_1}{2} & \text{if } x = 0 \\
        \alpha_1 e^{-ikx} & \text{if } 0 < x < d \\
        \frac{\alpha_1}{2} e^{-ikd} & \text{if } x = d \\
        0 & \text{if } x > d \end{cases} \\
    & \varphi_{R1}(x)  = \begin{cases} e^{ikx} & \text{if } x < 0 \\
        \frac{1 + \beta_1}{2} & \text{if } x = 0 \\
        \beta_1 e^{ikx} & \text{if } 0 < x < d \\
        \frac{\beta_1 + t_1}{2} e^{ikd} & \text{if } x = d \\
        t_1 e^{ikx} & \text{if } x > d \end{cases} \\ \nonumber
\end{align}
\end{subequations}
Here $\alpha_1$, $\alpha_2$, $\beta_1$, and $\beta_2$ are transport coefficients for the region between the coupling points, with $\beta$ and $\alpha$ referring to the right and left directions, respectively. By solving for the transport coefficients at the end of the ports, $t_q$ and $r_q$ for $q=1,2$, we obtain the following results:
\begin{subequations}
\begin{align}
    & t_1  = \frac{2\Gamma\big(1+e^{i\theta}+i\sin\theta\big) - i\Delta}{4\Gamma(1+e^{i\theta})-i\Delta},~r_1  = -\frac{\Gamma(1+e^{i\theta})^2}{4\Gamma(1+e^{i\theta})-i\Delta}, \\
    & t_2 = -\frac{\Gamma(1+e^{i\theta})^2e^{-i\theta}}{4\Gamma(1+e^{i\theta})-i\Delta},~\text{and}~r_2 = ~-\frac{\Gamma(1+e^{i\theta})^2}{4\Gamma(1+e^{i\theta})-i\Delta}
    ,
\end{align}    
\end{subequations}
where the phase accumulation between consecutive coupling points $\theta$ is given by $\theta = kd$ and the atom-waveguide coupling rate (assumed to be the same for both waveguides) $\Gamma\equiv V^2/v_g$. These transport coefficients directly relate to the transmission and reflection probabilities $T_q$ and $R_q$ by $T_q=|t_q|^2$ and $R_q=|r_q|^2$. 

These equations give us the reflection and transmission probability profiles shown in Figure~\ref {singleAtomPlots}. Several key physical features emerge from these plots. First, in the small-atom limit ($\theta \to 0$), the transmission and reflection amplitudes simplify to:
\begin{subequations}
\begin{align}
    t_1 &= \frac{4\Gamma - i\Delta}{8\Gamma - i\Delta}, \\
    r_1 = t_2 = r_2 &= -\frac{4\Gamma}{8\Gamma - i\Delta}.
\end{align}
\end{subequations}
We note that these equations reduce precisely to the multi-channel waveguide QED results for a single point-like atom previously studied (see, for example, Ref.~\cite{poudyal2020collective}). The effective decay rate into each waveguide channel becomes $\Gamma_{\text{chnl}} = 4\Gamma$, yielding a total collective decay rate of $\Gamma_{\text{tot}} = 8\Gamma$ due to the constructive interference of the overlapping coupling points. Secondly, we notice that when $\theta$ is an odd multiple of $\pi$ (for example, at $\theta = \pi$), the atom completely decouples from both waveguides, allowing the photon to transmit through the first waveguide. Furthermore, the transport properties show a symmetry under $x \to -x$ in waveguide 2. This behavior stems directly from the system's underlying geometry: Waveguide 2 keeps its structural inversion symmetry, while Waveguide 1 breaks it due to the incident photon's directional bias. These distinct symmetric and asymmetric features provide an important baseline, since they either remain or vanish in the various multi-atom configurations analyzed later. In the case of a single giant atom with two coupling points positioned at symmetric distances, we note that there is only one configuration possible, which has been previously examined in Ref.~\cite{wang2021giant}. Finally, the routing efficiency into the secondary channel, in both the forward and backward directions, is quantified using the metric $\max\{T_2(\Delta,\theta) + R_2(\Delta,\theta)\}$. In this single-atom case, we find that the efficiency cannot exceed $0.5$. As a result, this basic setup is limited and inefficient for single-photon routing applications, motivating the need for the multi-atom arrays considered in subsequent sections.


\section{\label{sec:III} Two giant atoms}
\begin{figure}
    \centering
    \begin{tabular}{rr}
        \text{(a)} & \includegraphics[width=2.7in, height=2in]{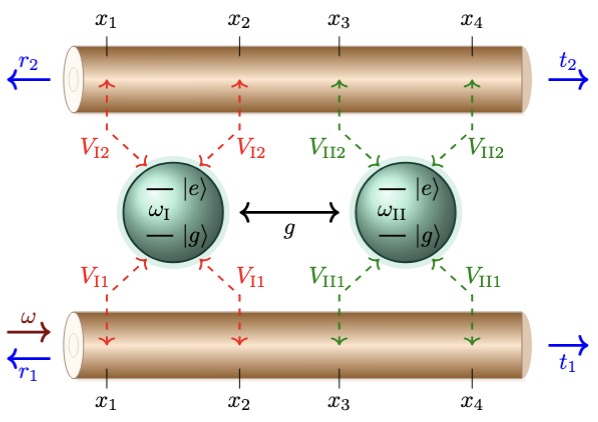} \\
        \noalign{\vspace{0.65cm}} 
        
        \text{(b)} & \includegraphics[width=2.7in, height=2in]{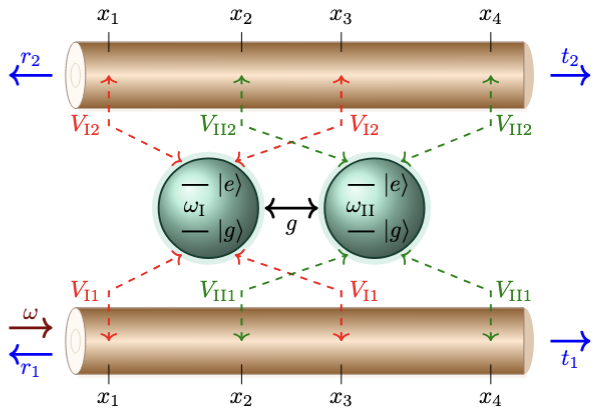} \\
        \noalign{\vspace{0.65cm}}
        
        \text{(c)} & \includegraphics[width=2.7in, height=3in]{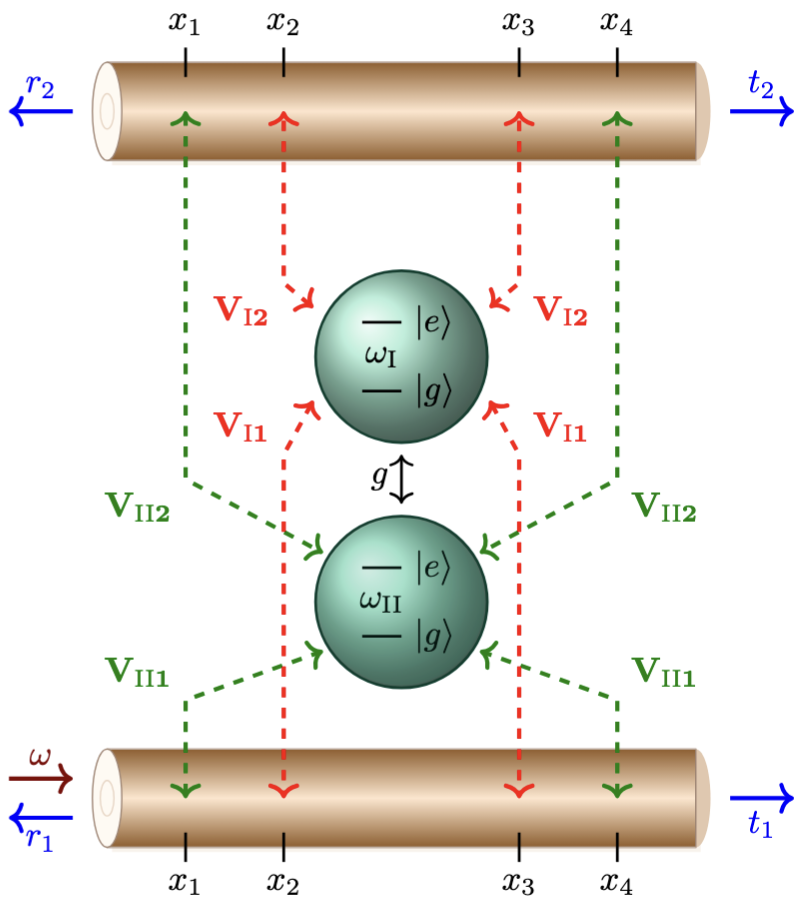} \\
    \end{tabular}
    \captionsetup{
format=plain,
margin=1em,
justification=raggedright,
singlelinecheck=false
}
    \caption{(Color online) A single photon router using two giant atoms, coupled to a top and bottom waveguide at points $ x_1, x_2, x_3,$ and $x_4$ with coupling strength parameters $V_{\text{I}1}$, $V_{\text{I}2}$, $V_{\text{II}1}$, and $V_{\text{II}2}$ in a (a) fully separated, (b) simply braided, and (c) fully nested configuration.}
    \label{2Gaint}
\end{figure}

Introducing a second giant atom creates multiple topologically distinct arrangements for assigning coupling points along the waveguides. In a two-waveguide framework, we focus on symmetric layouts in which both waveguides share identical coupling topologies \cite{sun2025frequency}. Hereafter, we refer to these arrangements as {\it fully separate}, {\it simply braided}, and {\it fully nested}, respectively (see Fig.~\ref{2Gaint}).

The generic form of Hamiltonian for two giant atom routing problems applicable to any of the aforementioned topologies is given by (with $\hbar = 1$): 
\begin{align}
    &\hat{\mathcal{H}} = \sum^{\text{II}}_{j = \text{I}} \omega_j \hat{\sigma}_j^\dagger \hat{\sigma}_j -i v_g \sum^2_{q=1} \int_{-\infty}^{+\infty} dx \bigg[ \hat{c}^\dagger_{Rq}(x) \partial_x \hat{c}_{Rq}(x)\nonumber\\
    &- \hat{c}_{Lq}^\dagger(x) \partial_x \hat{c}_{Lq}(x)\bigg] + \sum_{p \in \{L,R\}} \sum^2_{q = 1} \sum^{\text{II}}_{j = \text{I}} \sum_{l=1}^4 V C_{ql}^j\times\nonumber\\
    &\left[ \hat{c}_{pq}^\dagger(x_{ql}) \hat{\sigma}_j + \hat{\sigma}^\dagger_j\hat{c}_{pq}(x_{ql}) \right] + g \left[ \hat{\sigma}_{\text{I}}^\dagger \hat{\sigma}_{\text{II}} + \hat{\sigma}_{\text{II}}^\dagger \hat{\sigma}_{\text{I}} \right].
\end{align}
Here, $\hat{\sigma}_j^\dagger$ ($\hat{\sigma}_j$) is the raising (lowering) operator for the $j^{\text{th}}$ atom, and $g$ is the direct, non-waveguide-mediated coupling strength between the two atoms. The photon ladder operators $\hat{c}_{pq}^\dagger$ ($\hat{c}_{pq}$) and the uniform waveguide group velocity $v_g$ are defined as before. Finally, the configuration-dependent term $C_{ql}^j$ equals $1$ if atom $j$ couples to point $x_{ql}$, and equals $0$ otherwise.

We now extend the single-photon excitation state to the case of two giant atoms ($j = \text{I}, \text{II}$). The net state vector is modified to include both atomic probability amplitudes $u_j$:
\begin{align}
    |\psi\rangle = &  \bigg(\sum^{\text{II}}_{j = \text{I}}u_j\hat{\sigma}_j^\dagger + \sum_{p \in \{L,R\}} \sum^2_{q=1} \int^{+\infty}_{-\infty} dx \, \varphi_{pq}(x) \hat{c}_{pq}^\dagger (x)\bigg) |\varnothing\rangle,
\end{align}
where $u_j$ now accounts for the amplitude of either atom $\text{I}$ or $\text{II}$ to be excited with no photons in the waveguides, and $\varphi_{pq}(x)$ remains the waveguide probability amplitudes and $\ket{\varnothing}$ the global ground state as before. With two distinct atoms, the atomic detuning generalizes to $\Delta_j = \omega - \omega_j$, while the photon wavenumber $k = \omega/v_g$ and the coupling-dependent decay rates $\Gamma_{pql} = V_{pql}^2/v_g$ are defined analogously to the single giant atom case.

After applying the appropriate projections to the time-independent Schr\"{o}dinger equation, we obtain the following sets of equations obeyed by the probability amplitudes
\begin{subequations}
\begin{align}
    & u_j\Delta_j - \sum_{p=L,R} \sum^2_{q = 1} \sum_{l=1}^4 \left[u_k g + V_{pql} C_{ql}^j \varphi_{pq}(x_{ql})\right]=0, \label{atom} \\
    & \frac{d \varphi_{Lq}(x)}{d x} = -ik\varphi_{Lq}(x) +\sum^\text{II}_{j = \text{I}} \sum_{l=1}^4 C_{ql}^j\left[ \frac{i\Gamma_{Lql}}{V_{Lql}} u_j \delta(x- x_{ql}) \right], \label{jumpL} \\
    & \frac{d \varphi_{Rq}(x)}{d x}  =ik\varphi_{Rq}(x) - \sum^{\text{II}}_{j = \text{I}} \sum_{l=1}^4 C_{ql}^j \left[\frac{i\Gamma_{Rql}}{V_{Rql}} u_j \delta(x-x_{ql}) \right]. \label{jumpR}
\end{align}
\end{subequations}
Here again $j=\text{I},\text{II}$, with $q=1,2$. Next, proceeding as in the single giant atom case, we apply the Bethe ansatz to the two-giant-atom routing system. Enforcing the regularization condition at every point of discontinuity and setting $x_1=0$ allows us to write:
{\allowdisplaybreaks
\begin{subequations}
    \begin{align}
       &\varphi_{L2}(x) = \begin{cases} 
        r_2 e^{-ikx} & \text{if } x < x_1 \\
        \frac{r_2 + \alpha_{21}}{2} & \text{if } x = x_1 \\
        \alpha_{21} e^{-ikx} & \text{if } x_1 < x < x_2 \\
        \frac{\alpha_{21}+\alpha_{22}}{2} e^{-ikx_2} & \text{if } x = x_2 \\
        \alpha_{22}e^{-ikx} & \text{if } x_2 < x < x_3 \\
        \frac{\alpha_{22}+\alpha_{23}}{2} e^{-ikx_3} & \text{if } x = x_3 \\
        \alpha_{23}e^{-ikx} & \text{if } x_3 < x < x_4 \\
        \frac{\alpha_{23}}{2} e^{-ikx_4} & \text{if } x = x_4 \\
        0 & \text{if } x > x_4 \\
        \end{cases} \\
    & \varphi_{R2}(x) = \begin{cases} 
        0 & \text{if } x < x_1 \\
        \frac{\beta_{21}}{2} & \text{if } x = x_1 \\
        \beta_{21} e^{ikx} & \text{if } x_1 < x < x_2 \\
        \frac{\beta_{21}+\beta_{22}}{2} e^{ikx_2} & \text{if } x = x_2 \\
        \beta_{22}e^{ikx} & \text{if } x_2 < x < x_3 \\
        \frac{\beta_{22}+\beta_{23}}{2} e^{ikx_3} & \text{if } x = x_3 \\
        \beta_{23}e^{ikx} & \text{if } x_3 < x < x_4 \\
        \frac{\beta_{23} + t_2}{2} e^{ikx_4} & \text{if } x = x_4 \\
        t_2 e^{ikx} & \text{if } x > x_4 \\
        \end{cases} \\
    &\varphi_{L1}(x) = \begin{cases} 
        r_1 e^{-ikx} & \text{if } x < x_1 \\
        \frac{r_1 + \alpha_{11}}{2} & \text{if } x = x_1 \\
        \alpha_{11} e^{-ikx} & \text{if } x_1 < x < x_2 \\
        \frac{\alpha_{11}+\alpha_{12}}{2} e^{-ikx_2} & \text{if } x = x_2 \\
        \alpha_{12}e^{-ikx} & \text{if } x_2 < x < x_3 \\
        \frac{\alpha_{12}+\alpha_{13}}{2} e^{-ikx_3} & \text{if } x = x_3 \\
        \alpha_{13}e^{-ikx} & \text{if } x_3 < x < x_4 \\
        \frac{\alpha_{13}}{2} e^{-ikx_4} & \text{if } x = x_4 \\
        0 & \text{if } x > x_4 \\
        \end{cases} \\
    & \varphi_{R1}(x) = \begin{cases} 
        e^{ikx} & \text{if } x < x_1 \\
        \frac{1 + \beta_{11}}{2} & \text{if } x = x_1 \\
        \beta_{11} e^{ikx} & \text{if } x_1 < x < x_2 \\
        \frac{\beta_{11}+\beta_{12}}{2} e^{ikx_2} & \text{if } x = x_2 \\
        \beta_{12}e^{ikx} & \text{if } x_2 < x < x_3 \\
        \frac{\beta_{12}+\beta_{13}}{2} e^{ikx_3} & \text{if } x = x_3 \\
        \beta_{13}e^{ikx} & \text{if } x_3 < x < x_4 \\
        \frac{\beta_{13} + t_1}{2} e^{ikx_4} & \text{if } x = x_4 \\
        t_1 e^{ikx} & \text{if } x > x_4 \\ \end{cases}
    \end{align}
\end{subequations}}
Here, $\alpha_{ql}$ ($\beta_{ql}$) is the probability amplitude for finding a photon in the left (right) propagating mode within the region $x_{ql}<x<x_{q(l+1)}$. The terms $r_q$ and $t_q$ represent the probability amplitudes for a photon being reflected or transmitted in waveguide $q$. Applying the jump conditions from Eq.~\eqref{jumpL} and Eq.~\eqref{jumpR}, we obtain:
\begin{align}
   C_{ql}^j \frac{i\Gamma_{Lql}}{V_{Lql}}u_j & = C_{ql}^j \times\begin{cases} 
        (\alpha_{q1} - r_q) e^{-ikx_{q1}}, & l=1 \\
        (\alpha_{ql} - \alpha_{q(l-1)})e^{-ikx_{ql}}, & l=2,3 \\
        -\alpha_{q3} e^{-ikx_{q4}}, & l=4
   \end{cases} \\[1ex]
  - C_{ql}^j \frac{i\Gamma_{Rql}}{V_{Rql}}u_j & = C_{ql}^j \times \begin{cases} 
        (\beta_{q1}-\delta_{q1}) e^{ikx_{q1}}, & l=1 \\
        (\beta_{ql} - \beta_{q(l-1)})e^{-ikx_{ql}}, & l=2,3 \\
        (t_q-\beta_{q3}) e^{ikx_{q4}}, & l=4
   \end{cases}    
\end{align}
where $\delta_{q1}$ is the Kronecker delta (which in our notation is equal to $1$ for waveguide $q=1$, and $0$ for $q=2$). Substituting the photon amplitudes into Eq.~\eqref{atom} yields the following two coupled equations for each atom $j=\text{I}, \text{II}$ (with $k' \neq j$):
\begin{align}
    &u_j \Delta_j - u_{k'} g - \sum_{q=1,2}\sum_{l=1}^{4}\frac{V_{pql}}{2} C_{ql}^{j} \bigg[ (\alpha_{ql} + \alpha_{q(l-1)})e^{-ikx_{ql}}\nonumber\\
    &+ (\beta_{ql} + \beta_{q(l-1)})e^{ikx_{ql}} \bigg] =0,
\end{align}
subject to the boundary conditions $\alpha_{q0}=r_q$, $\alpha_{q4}=0$, $\beta_{q0}=\delta_{q1}$, and $\beta_{q4}=t_q$. Note that while the single giant-atom system yields simple analytical solutions, the two-atom equations lead to highly involved expressions for the reflection and transmission coefficients. Consequently, we solve these equations numerically and present the results graphically below.

\begin{figure*}
\includegraphics[width=6.8in, height=1.7in]{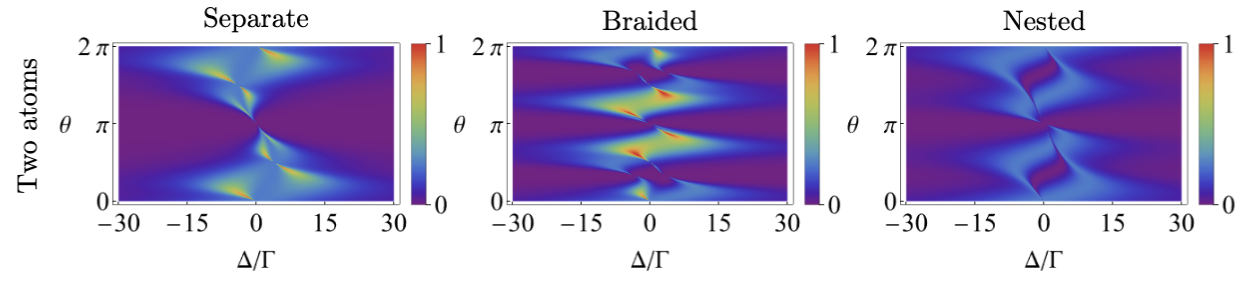} 
\captionsetup{
format=plain,
margin=1em,
justification=raggedright,
singlelinecheck=false
}
\caption{(Color online) The routing efficiency $|t_2|^2$ as a function of the normalized detuning $\Delta/\Gamma$ and the accumulated phase $\theta$, assuming identical atoms ($\Delta_{\text{I}} = \Delta_{\text{II}} = \Delta$), uniform decay rates ($\Gamma_{pql} = \Gamma$), and vanishing direct coupling ($g/\Gamma = 0$). For the separate and nested configurations, the coupling points are equally spaced such that the phase accumulation between any two adjacent coupling points is given by $\theta$. For the braided configuration, the phase accumulation is non-uniform, satisfying $2\theta = k(x_2 - x_1) = k(x_4 - x_3)$ and $\theta = k(x_3 - x_2)$.}
\label{fig:2AtomBasic}
\end{figure*}
\begin{figure*}
\includegraphics[width=6.8in, height=1.7in]{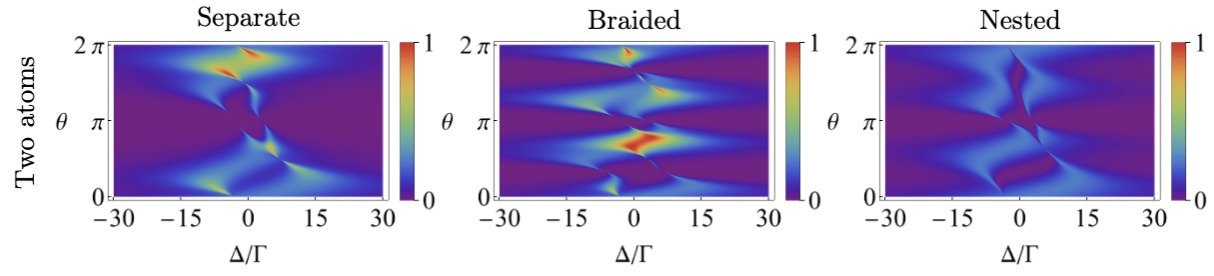} 
\captionsetup{
format=plain,
margin=1em,
justification=raggedright,
singlelinecheck=false
}
\caption{(Color online) Routing efficiency $|t_2|^2$ is shown as a function of normalized detuning $\Delta/\Gamma$ and accumulated phase $\theta$ for a non-zero direct interaction strength $g/\Gamma = 3$. All other parameters and coupling configurations are identical to those in Fig.~\ref{fig:2AtomBasic}.}
\label{fig:2AtomCoupling}
\end{figure*}
As shown in Fig.~\ref{fig:2AtomBasic}, the routing efficiency $|t_2|^2$ into the second waveguide reveals that for the fully separate configuration, high-efficiency routing ($|t_2|^2 \to 1$) is tightly restricted to a narrow frequency band around the atomic resonance ($\Delta/\Gamma = 0$) and localized phase values ($\theta$). On the contrary, the simply braided arrangement opens non-local interference channels, which yield broad, periodic bands of perfect transmission that are robust against variations in photon detuning. Finally, the fully nested configuration severely limits the parameter space for efficient photon transfer, capping the maximum routing probability at exactly $0.25$. This strict limitation arises from the distinct spatial inversion symmetry of the nested topology. Unlike the separate and braided configurations (where spatial inversion permutes the relative ordering of the atoms), the nested configuration remains entirely invariant under coordinate inversion. Consequently, an absorbed photon cannot distinguish between left-moving and right-moving directions within the waveguides, forcing a symmetric distribution of the scattered field ($T_{2} = R_{2} = R_{1} = T_{1} = 0.25$). This causes the nested configuration to fundamentally mirror the routing bounds of a single giant atom. These results demonstrate that a braided topology is an efficient, wideband directional router across all three topologies.

The presence of a second giant atom opens the possibility of direct or non-waveguide-mediated atom-atom interactions \cite{kockum2018decoherence, wang2021giant}. In superconducting quantum circuits, such direct coupling $g$ may arise from near-field electrostatic capacitance between qubit pads, or mutual inductance between loop structures. In Fig.~\ref{fig:2AtomCoupling} we report the transmission profile $|t_2|^2$ for each of the three configurations with a non-zero interatomic coupling strength ($g \neq 0$). We note that the points of effective decoupling (characterized by vanishing routing efficiency ($|t_2|^2 \to 0$) at specific values of the phase accumulation $\theta$) persist despite the direct interaction. However, this direct coupling explicitly breaks the generalized inversion symmetry, $|t_2(\theta, \Delta)|^2 = |t_2(-\theta, -\Delta)|^2$, that characterized the $g=0$ regime (see Fig.~\ref{fig:2AtomBasic} for comparison). We also note that although the fully separated and simply braided configurations still preserve regions of high routing efficiency, the fully nested configuration remains highly suppressed, with a maximum routing probability of only $0.25$.


\section{\label{sec:IV} Extension to many giant atoms}
\begin{figure}
    \centering
    \begin{tabular}{rr}
        \text{(a)} & \includegraphics[width=3.3in, height=1.4in]{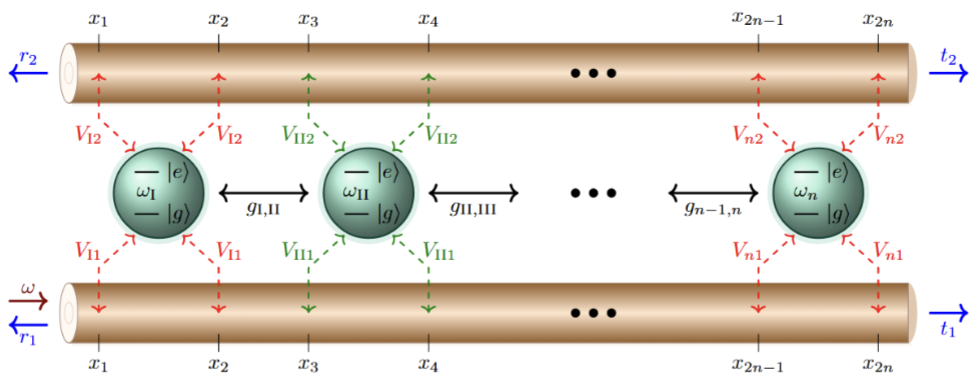} \\
        \noalign{\vspace{0.5cm}} 
        
        \text{(b)} & \includegraphics[width=3.3in, height=1.4in]{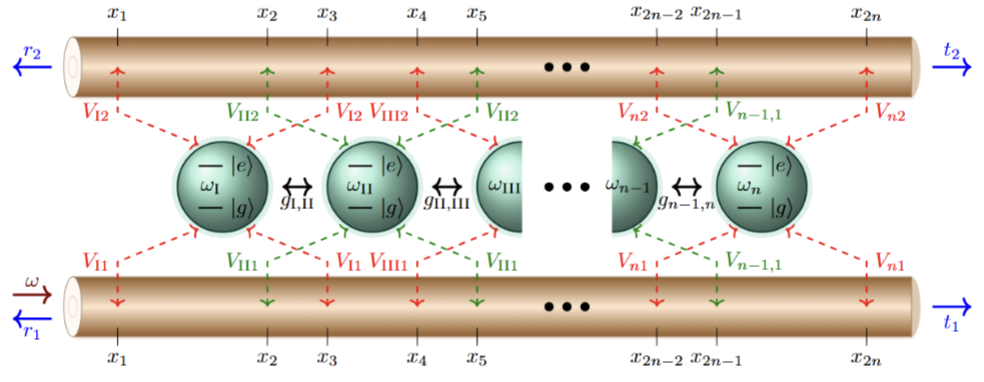} \\
        \noalign{\vspace{0.5cm}}
        
        \text{(c)} & \includegraphics[width=3.3in, height=2.5in]{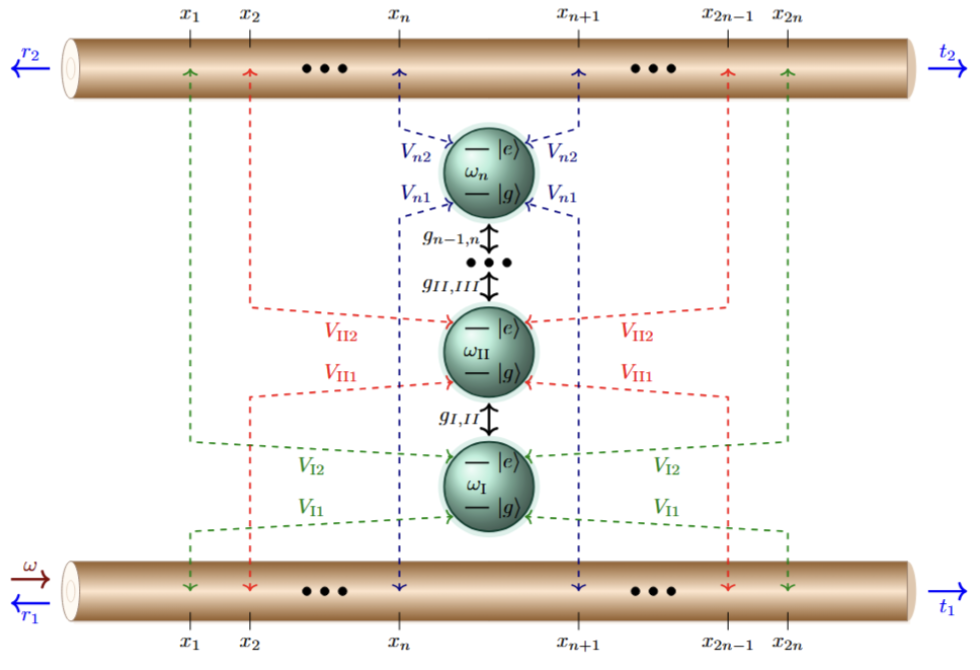} \\
    \end{tabular}
    \captionsetup{
format=plain,
margin=1em,
justification=raggedright,
singlelinecheck=false
}
    \caption{(Color online) Schematic representations of a single-photon router mediated by multiple giant atoms under three topological configurations: (a) fully separated, (b) simply braided, and (c) fully nested. As before, the system consists of two parallel waveguides coupled via $n$ giant atoms, incorporating a nearest-neighbor direct interatomic interaction $g_{j,j+1}$. The discrete coupling nodes are located at coordinates $x_1, \dots, x_{2n}$, where the relative spatial separation between adjacent connection points is assumed to be identical for both waveguides.}\label{fig:models}
\end{figure}
We now generalize our framework to a system consisting of $n$ giant atoms coupled simultaneously to two linear waveguides (indexed as $q=1,2$). Each atom $j \in \{\text{I}, \text{II}, \dots, n\}$ is modeled as a two-level system with ground state $|g\rangle$, excited state $|e\rangle$, and transition frequency $\omega_j$. Extending the two-point coupling geometry discussed previously, each of the $n$ giant atoms interacts with both waveguides at two discrete spatial locations chosen from the ordered coordinate set $x_l$ (where $l \in \{1, 2, \dots, 2n\}$). For simplicity, we assume a uniform coupling amplitude $V_{jq}$ for atom $j$ across both waveguides. The three topological configurations considered in this work for these multi-atom layouts are illustrated schematically in Fig.~\ref{fig:models}.

Building upon the single- and two-giant atom models introduced in the previous sections, we now generalize the system to $n$ giant atoms. Utilizing a real-space formalism \cite{shen2005coherent, shen2005coherent1} to describe the continuous waveguide fields and localized atom-waveguide interactions, and extending the multi-point coupling frameworks of \cite{wang2021giant, sun2025frequency} to an arbitrary number of giant atoms, the net Hamiltonian of the system is given by
\begin{align}
    &\hat{\mathcal{H}} = \sum_{j=\text{I}}^{n} \omega_j \hat{\sigma}_j^\dagger\hat{\sigma}_j + \sum_{j=I}^{n} \sum_{k'=j+1}^{n} g_{j,k'} \left( \hat{\sigma}_j^\dagger \hat{\sigma}_{k'} + \hat{\sigma}_{k'}^{\dagger} \hat{\sigma}_j \right) \nonumber \\
    &  - i v_g \sum^2_{q=1} \int dx \left[ \hat{c}_{Rq}^\dagger(x) \partial_x \hat{c}_{Rq}(x) - \hat{c}_{Lq}^\dagger (x) {\partial_x} \hat{c}_{Lq}(x) \right]\nonumber \\
    & + \sum_{p \in \{L,R\}} \sum^2_{q=1} \sum_{j=\text{I}}^{n} \sum_{l=1}^{2n} VC_{ql}^j \left[ \hat{c}_{pq}^\dagger(x_{ql})\hat{\sigma}_{j} + \hat{\sigma}_{j}^\dagger\hat{c}_{pq}(x_{ql}) \right],
\end{align}
where $\hat{\sigma}_j^\dagger$ ($\hat{\sigma}_j$) denotes the atomic raising (lowering) operator for the $j^{\text{th}}$ atom, and $g_{j,{k'}}$ represents the direct, non-waveguide-mediated coupling strength between atoms $j$ and $k'$. The operators $\hat{c}_{pq}^\dagger(x)$ and $\hat{c}_{pq}(x)$ and $v_g$ being the group velocity are defined as before. The coordinate $x_{ql}$ specifies the spatial position of the $l^{\text{th}}$ coupling point ($l \in \{1,2\}$) along waveguide $q$ with $C_{ql}^j$ taking into account the $j$th atom. 

Like before, we consider a scattering scenario in which a single photon with frequency $\omega$ is incident from the left in waveguide 1. Within the single-excitation subspace, the eigenstates of the system can be expanded as
\begin{equation}
    |\psi\rangle = \bigg(\sum_{j=\text{I}}^n u_j \hat{\sigma}_j^\dagger + \sum_{p \in \{L,R\}} \sum^2_{q=1} \int_{-\infty}^\infty dx \, \varphi_{pq}(x) \hat{c}_{pq}^\dagger (x) \bigg)|\varnothing \rangle,
\end{equation}
where $u_j$ is the probability amplitude of finding the $j^{\text{th}}$ giant atom in its excited state with no photons in the waveguide and all other atoms unexcited, and $\varphi_{pq}(x)$ represents the continuous real-space probability amplitude of a photon propagating in direction $p$ within waveguide $q$ with all atoms in their ground state. The state $|\varnothing\rangle$ denotes the global ground state, defined as before, corresponding to all $n$ giant atoms residing in their respective ground states and zero photons in either waveguide.

Proceeding further, as before, we apply the Bethe ansatz and adopt a piecewise plane-wave solution for the continuous waveguide fields within each spatial interval. In a compact notation, we thus write
\begin{subequations} \label{ansatz}
\begin{align}
    \varphi_{Lq}(x) &= \begin{cases} 
        r_q e^{-ikx}, & x < x_{q,1} \\ 
        \alpha_{q,l} e^{-ikx}, & x_{q,l} < x < x_{q,l+1} \\ 
        0, & x > x_{q,2n} 
    \end{cases} \\[1.5ex]
    \varphi_{Rq}(x) &= \begin{cases} 
        \delta_{q,1} e^{ikx}, & x < x_{q,1} \\ 
        \beta_{q,l} e^{ikx}, & x_{q,l} < x < x_{q,l+1} \\ 
        t_{q} e^{ikx}, & x > x_{q,2n} 
    \end{cases}
\end{align}
\end{subequations}
where $k = \omega/v_g$ represents the photon wavenumber, $l \in \{1, 2, \dots, 2n-1\}$, and $\delta_{q,1}$ is the Kronecker delta describing that the single-photon was launched into waveguide 1. The coefficients $r_q$ and $t_q$ denote the global reflection and transmission amplitudes for waveguide $q$, respectively. Next, regularizing the wavefunctions at the singular interaction points $x_{ql}$, we express:
\begin{subequations}\label{regularization}
\begin{align}
    \varphi_{Lq}(x_{q,l}) &= \frac{\alpha_{q,l-1} + \alpha_{q,l}}{2} e^{-ikx_{q,l}}, \\[1.5ex]
    \varphi_{Rq}(x_{q,l}) &= \frac{\beta_{q,l-1} + \beta_{q,l}}{2} e^{ikx_{q,l}}.
\end{align}
\end{subequations}
To maintain a unified notation across all intervals, we establish the boundary identifications $\alpha_{q,0} \equiv r_q$, $\alpha_{q,2n} \equiv 0$, $\beta_{q,0} \equiv \delta_{q,1}$, and $\beta_{q,2n} \equiv t_q$. Under this convention, the input condition $\beta_{1,0} = 1$ naturally satisfies the physical assumption that the single photon is incident from the far left ($x < x_{1,1}$) of the first waveguide.

By substituting Eq.~(\ref{ansatz}) and Eq.~(\ref{regularization}) into the Schr\"{o}dinger equation, we obtain the following system of probability amplitude equations, which can then be solved for transmission and reflection probabilities:
\begin{subequations}
    \begin{align}
        & u_j \Delta_j - \sum_{{k'}=\text{I}}^{n}u_{k'}g_{j{k'}} - \sum_{q=1}^{2}\sum_{l=1}^{2m}\frac{V_{jq}}{2} C_{ql}^{j} \big[ (\alpha_{ql} + 
    \alpha_{q(l-1)})e^{-ikx_{ql}}\nonumber\\
    &+ (\beta_{ql} + \beta_{q(l-1)})e^{ikx_{ql}} \big] = 0,\label{nAtomsEquation}\\
    & C_{ql}^{j} \left[ (\alpha_{ql} - \alpha_{q(l-1)})e^{-ikx_{ql}} - \frac{i\Gamma_{jq}}{V_{jq}}u_j\right] = 0,\label{nRightEquation}\\
    & C_{ql}^{j} \left[ (\beta_{ql} - \beta_{q(l-1)})e^{ikx_{ql}} + \frac{i\Gamma_{jq}}{V_{jq}}u_j\right] = 0,\label{nLeftEquation}
    \end{align}
\end{subequations}
where ${k'} \neq j$, $\Gamma_{jq} = \frac{V_{jq}^2}{v_g}$ and $C_{ql}^j = 1$ when atom $j$ couples to waveguide $q$ at point $x_{ql}$ and 0 otherwise, and is what determines what connection configuration we have between the giant atoms and waveguides. Eq.~(\ref{nAtomsEquation}) is repeated for each of the $n$ atoms, while Eqs.~(\ref{nRightEquation}) and (\ref{nLeftEquation}) repeat for each waveguide, for a total of $n + 2(2n) + 2(2n) =9n$ equations in the system (for each atom we add one atomic amplitude and four photon amplitudes for each direction). 


While this theoretical treatment holds for any number $n$ of giant atoms, an arbitrary number of coupling points per waveguide, and general spatial arrangements (including setups where multiple atoms share the same coupling nodes), we restrict the present study to cases where each atom couples to each waveguide at exactly two points. Even with this restriction, $n$ atoms yield $\left[ \frac{(2n)!}{(2!)^n n!} \right]^2$ possible coupling permutations. We focus on highly symmetric extensions of two-atom configurations: fully separate, simply braided, and fully nested (Fig.~\ref{fig:models}).

To simplify computations, we set equal distances between all consecutive coupling points in the separate and nested schemes. In the separate case, this ensures that all atoms are the same ``size'' (i.e., the distance between an atom's coupling points on a single waveguide). In the nested case, atomic size scales linearly with the nesting level. For the braided configuration, we prioritize equal atomic sizes, meaning the outermost coupling points on each waveguide are twice as far from their nearest neighbors as all other consecutive points (illustrated in Fig.~\ref{fig:models}(b)). Although alternative spacing schemes can optimize routing probabilities for different ports~\cite{sun2025frequency}, this specific arrangement effectively optimizes rightward transmission in the upper waveguide.

\subsection{The case of no direct atom-atom interaction}
\begin{figure*}
\centering
\includegraphics[width=6in, height=4in]{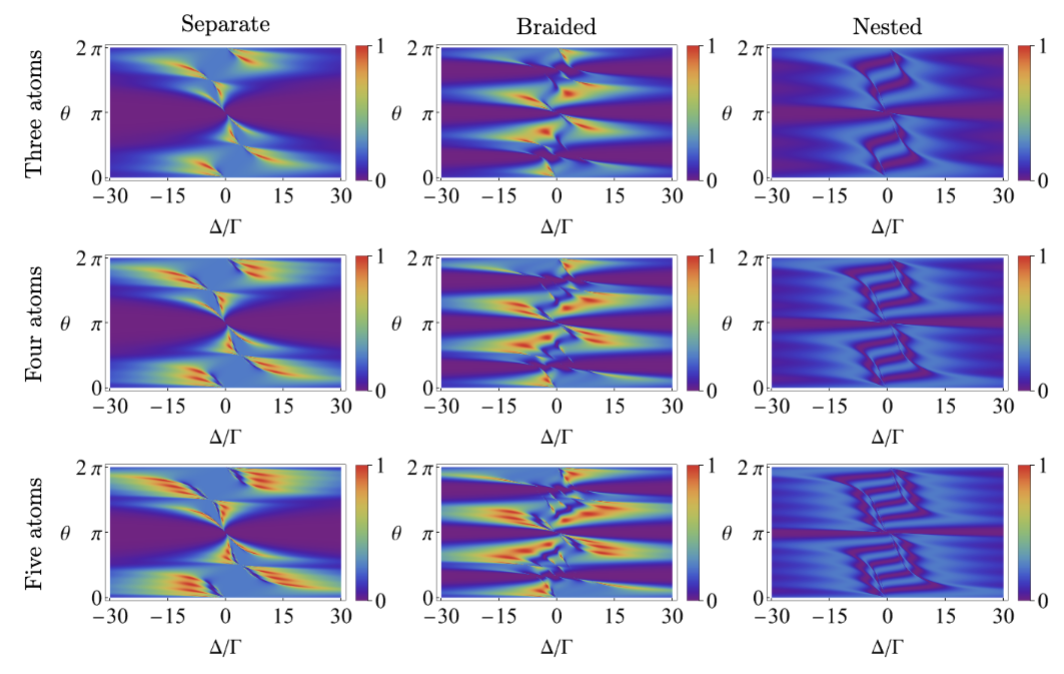}
    \captionsetup{
format=plain,
margin=1em,
justification=raggedright,
singlelinecheck=false
}
\caption{(Color online) Routing probability $|t_2|^2$ versus $\theta$ and $\Delta/\Gamma$ for identical $\Gamma_{jq}=\Gamma$ and $\Delta_j=\Delta$, with $g_{\text{all-to-all}}=0$. Separate and nested schemes have uniform $\theta$ spacing; the braided configuration has $2\theta$ outermost spacing and $\theta$ internal spacing.
}
\label{fig:forwardNoG}
\end{figure*}

Figure~\ref{fig:forwardNoG} shows the routing probability for systems with three, four, and five giant atoms as a function of the photon-atom detuning $\Delta/\Gamma$ and the inter-point phase accumulation $\theta$. Each plot is $2\pi$-periodic in $\theta$, where $\theta = 0$ corresponds to all atoms coupling to both waveguides at identical points. Similar to Fig.~\ref{fig:2AtomBasic}, points of effective decoupling (zero routing probability) occur regardless of the atom number. This happens at $\theta = \pi$ for all configurations, with additional decoupling points at $\theta = \pi/3$ and $5\pi/3$ in the braided case. Furthermore, all configurations retain the symmetry $|t_2(\theta,\Delta)|^2 = |t_2(-\theta,-\Delta)|^2$, visible as a rotational symmetry about the origin ($\theta = \Delta = 0$).

\begin{figure*}
\centering
\includegraphics[width=6in, height=4in]{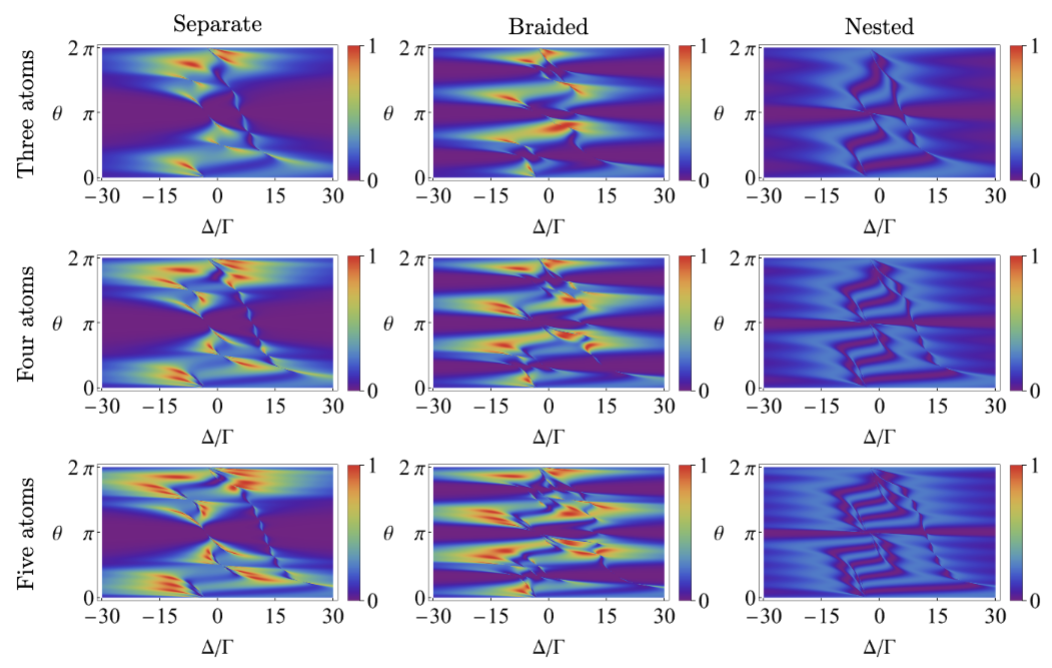}
    \captionsetup{
format=plain,
margin=1em,
justification=raggedright,
singlelinecheck=false
}
\caption{(Color online) Routing probability $|t_2|^2$ versus phase accumulation $\theta$ and detuning $\Delta/\Gamma$ with uniform all-to-all coupling $g_{\text{all-to-all}}=3\Gamma$. All other parameters and setup considerations are the same as Fig.~\ref{fig:forwardNoG}.}
\label{fig:manyA2A}
\end{figure*}

\begin{figure*}
\centering
\includegraphics[width=6in, height=4in]{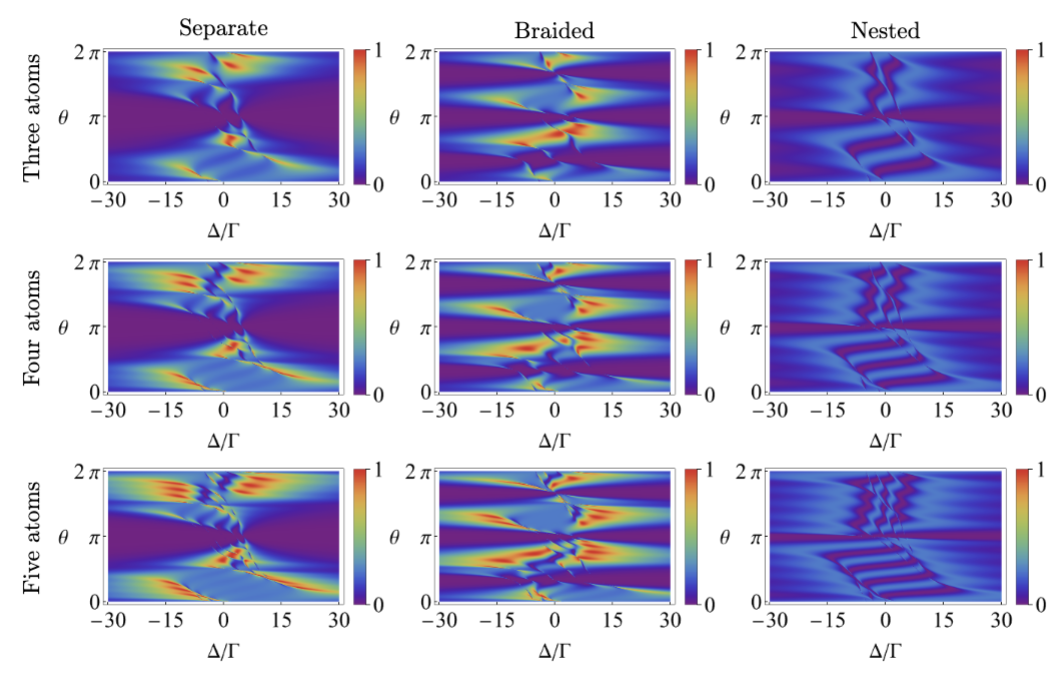}
    \captionsetup{
format=plain,
margin=1em,
justification=raggedright,
singlelinecheck=false
}
\caption{(Color online) Routing probability $|t_2|^2$ versus phase accumulation $\theta$ and detuning $\Delta/\Gamma$. All parameters and setup specifications are the  same as Fig.~\ref{fig:forwardNoG}, but with nearest-neighbor interaction $g_{\text{nn}} = 3\Gamma$ between consecutive atoms (and zero otherwise).}
\label{fig:manyNN}
\end{figure*}

In the separate and braided configurations, we note that increasing the number of giant atoms expands the parameter range for near-perfect routing. This advantage highlights the benefit of scaling up the single-photon router. Conversely, the maximum routing probability in the nested configuration remains capped at $0.25$. Because the nested setup preserves the single-atom symmetry, its $|r_1|^2$ and $|r_2|^2$ profiles are identical to $|t_2|^2$. In this regime, all photon emission directions are indistinguishable and equally likely, meaning the plots in Fig.~\ref{singleAtomPlots}, Fig.~\ref{fig:2AtomBasic}, and Fig.~\ref{fig:forwardNoG} merely reflect whether the photon is absorbed at all. Breaking this spatial symmetry (e.g., by varying relative coupling distances) provides a clearer path for steering routing directions.

\subsection{The case of direct atom-atom interactions}
Next, we examine the effect of atom-atom interactions. While a two-atom system permits only a single interatomic interaction, scaling to $n$ giant atoms introduces $\binom{n}{2}$ possible pairings. We consider two interaction topologies: uniform all-to-all coupling (i.e., $g_{\text{all-to-all}}\neq 0$), and nearest-neighbor-only coupling. These are modeled by setting $g_{\text{all-to-all}}/\Gamma = 3$ and $g_{\text{nn}}/\Gamma = 3$ for consecutive atoms (and zero otherwise). Figures~\ref{fig:manyA2A} and \ref{fig:manyNN} show the resulting routing probabilities for the all-to-all and nearest-neighbor coupling schemes, respectively.

From Fig.~\ref{fig:manyA2A} and Fig.~\ref{fig:manyNN}, we notice that in both interaction schemes, the points of effective decoupling remain invariant under the inclusion of interatomic interactions. However, the system's rotational symmetry is broken, such that $|t_2(\theta,\Delta)|^2 \neq |t_2(-\theta,-\Delta)|^2$. Scaling up the number of giant atoms continues to expand the parameter range for high-efficiency routing in both the separate and braided configurations, while sharpening the boundaries between high- and low-transmission regions. Conversely, the high spatial symmetry of the nested case remains detrimental, constraining its maximum routing probability to $0.25$. Notably, under all-to-all interactions, the separate and braided configurations exhibit a rigid shift in routing probability toward negative detunings. This interaction-induced detuning shift is entirely absent in the nearest-neighbor coupling scheme.

Beyond the global asymmetry, the structural differences between the two interaction topologies manifest as distinct spectral features. Under all-to-all coupling (Fig.~\ref{fig:manyA2A}), the uniform shift toward negative detunings suggests a global collective dressing of the atomic array, effectively acting as a cooperative Lamb shift that uniformly translates the system's resonance energy \cite{scully2009collective, rohlsberger2010collective}. Conversely, the nearest-neighbor scheme (Fig.~\ref{fig:manyNN}) reveals distinct multi-band splittings into discrete resonance branches, which become increasingly intricate as the number of atoms grows from three to five. Physically, because nearest-neighbor coupling respects the linear geometry of the chain rather than coupling all elements symmetrically, it gives rise to a discrete spectrum of collective eigenstates, leading to multiple split transmission pathways. Additionally, sharpening of the transmission boundaries with increasing $n$ in both cases suggests enhanced cooperative subradiant or superradiant interference effects, which strongly restrict or enhance photon transport within highly selective parameter windows \cite{poudyal2020collective, sheremet2023waveguide}.

\section{\label{sec:V} Inclusion of Losses}
\begin{figure*}
    \centering
    \begin{minipage}[t]{2.25in}
        \centering\textbf{(a)} \\ \vfill
        \includegraphics[width=2.25in, height=1.4in]{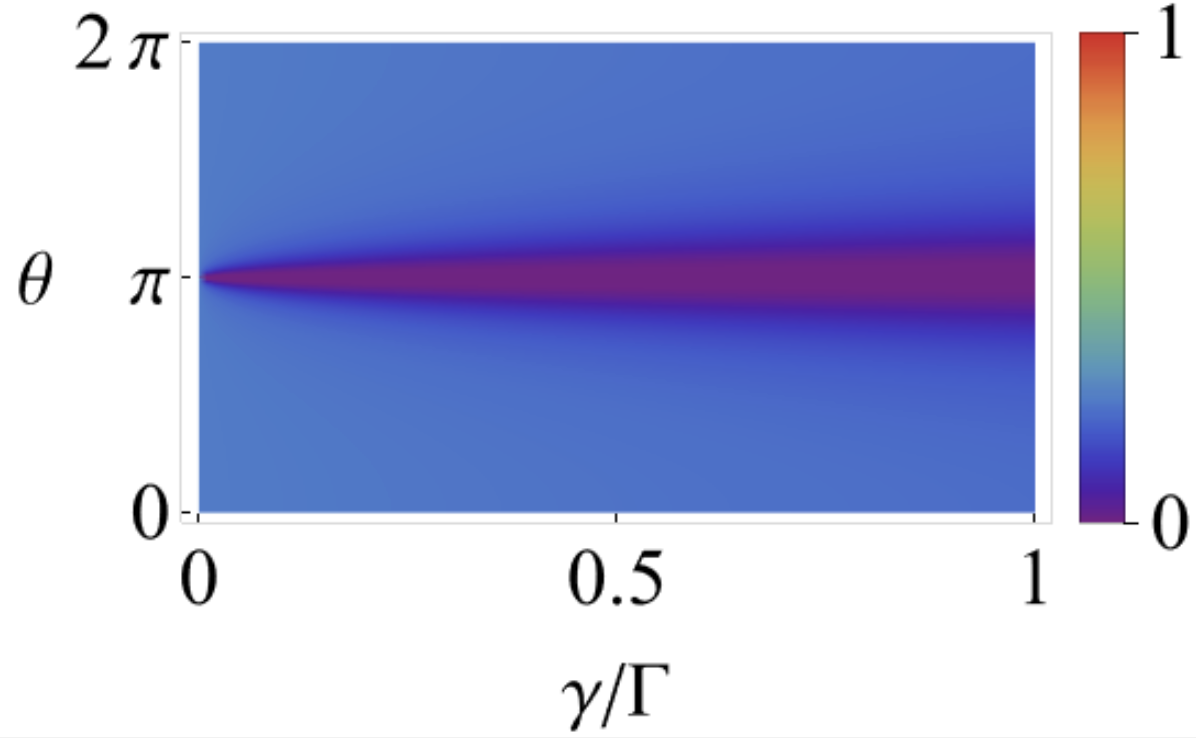}
    \end{minipage}\hspace{0.15in}
    \begin{minipage}[t]{2.25in}
        \centering\textbf{(b)} \\ \vfill
        \includegraphics[width=2.25in, height=1.4in]{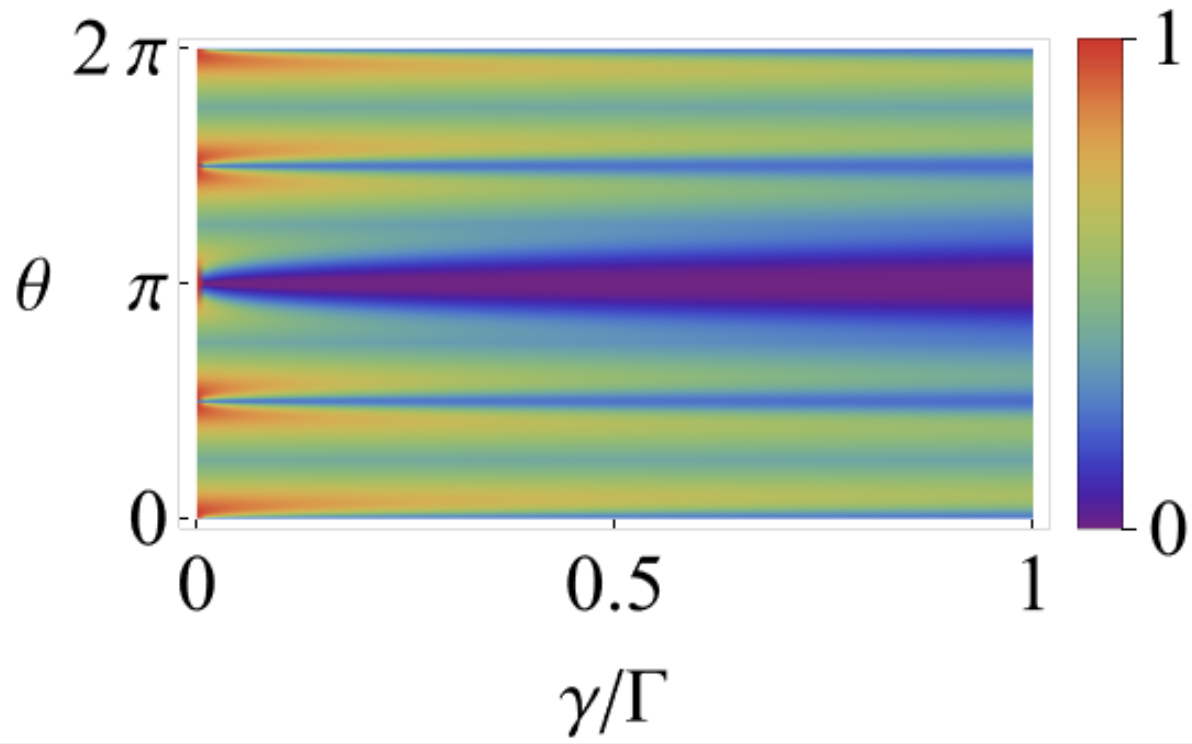}
    \end{minipage}\hspace{0.15in}
    \begin{minipage}[t]{2.25in}
        \centering\textbf{(c)} \\ \vfill
        \includegraphics[width=2.25in, height=1.4in]{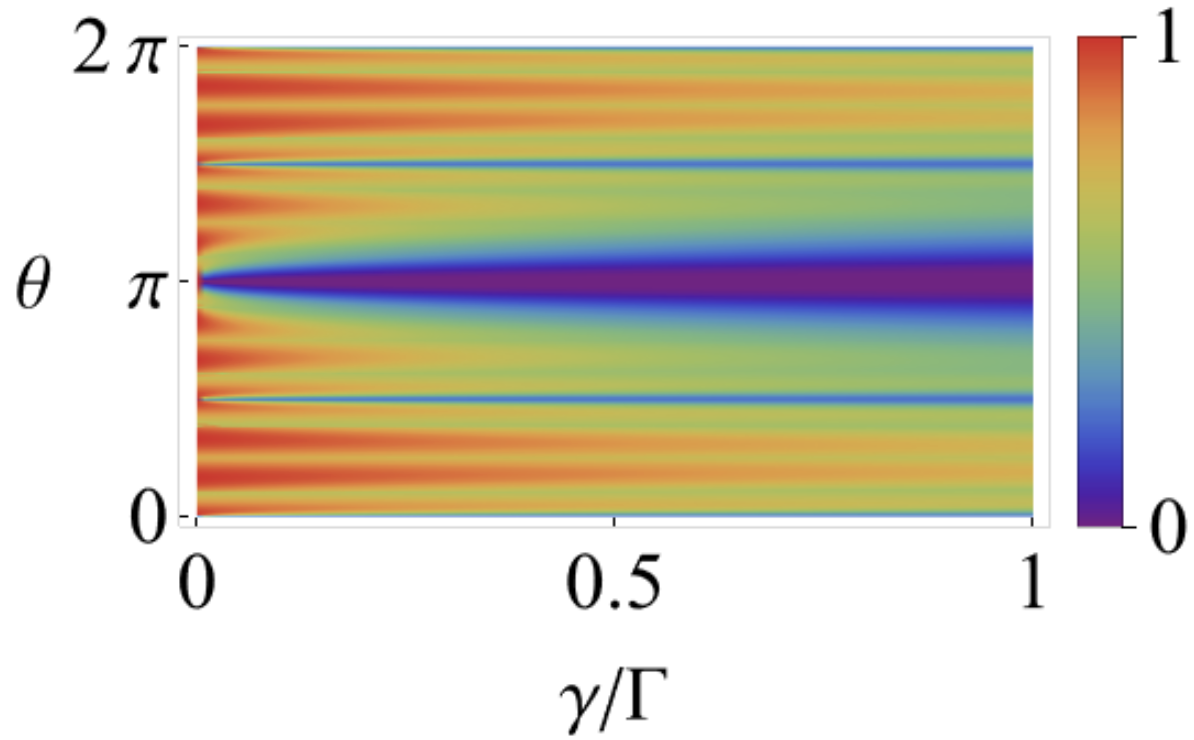}
    \end{minipage}

    \vspace{0.1in} 

    \begin{minipage}[t]{2.25in}
        \centering\textbf{(d)} \\ \vfill
        \includegraphics[width=2.25in, height=1.4in]{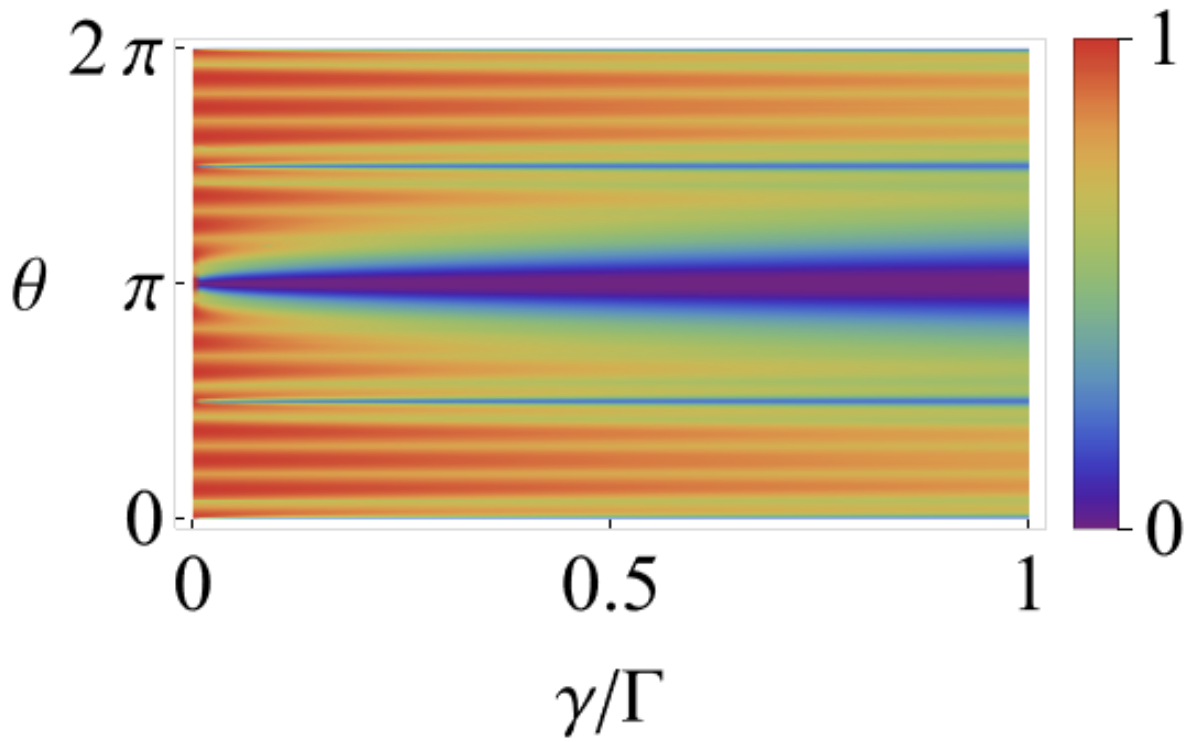}
    \end{minipage}\hspace{0.15in}
    \begin{minipage}[t]{2.25in}
        \centering\textbf{(e)} \\ \vfill
        \includegraphics[width=2.25in, height=1.4in]{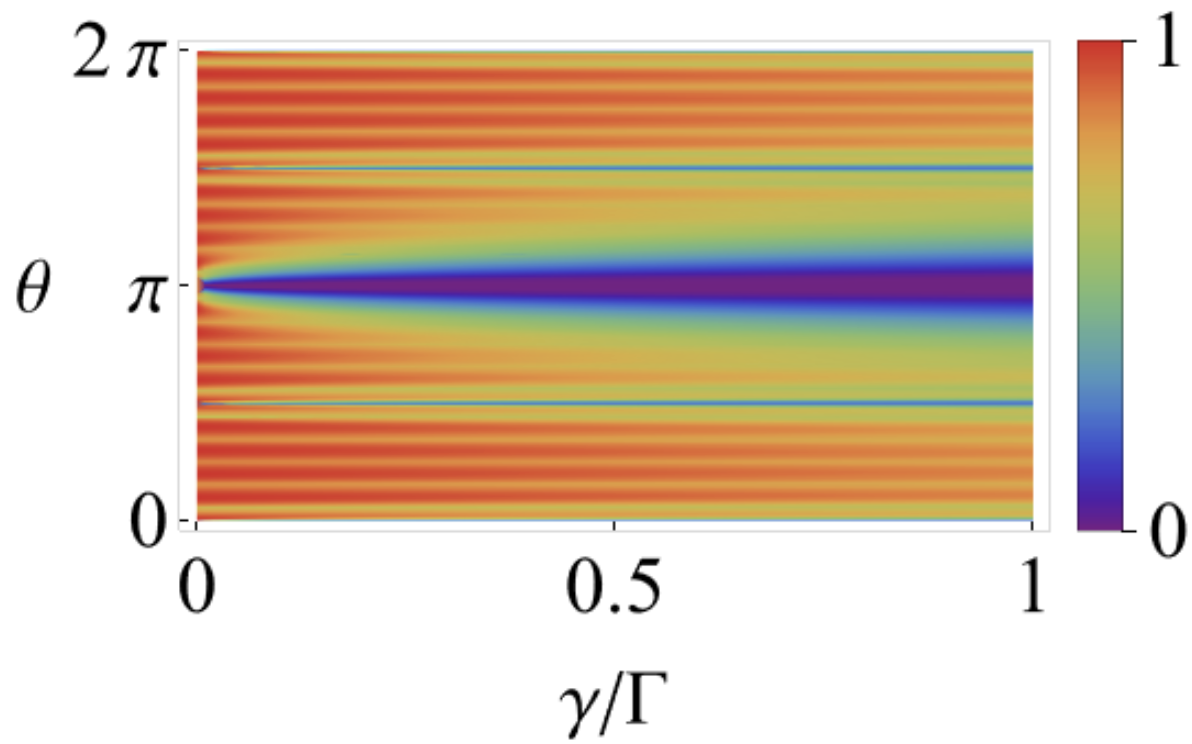}
    \end{minipage}
\captionsetup{
format=plain,
margin=1em,
justification=raggedright,
singlelinecheck=false
}
    \caption{(Color online) Maximum routing probability $\max_{\Delta} |t_2|^2$ into the forward-directed second waveguide as a function of the inter-point phase accumulation $\theta$ and the spontaneous emission rate $\gamma/\Gamma$. The optimization spans the detuning range $-30 \le \Delta/\Gamma \le 30$. All giant atoms are assumed identical with uniform waveguide coupling ($\Gamma_j = \Gamma$) and no direct coupling between giant atoms. Sub-figures (a), (b), (c), (d), and (e) show the maximum routing probability for one, two, three, four, and five giant atom chains, respectively, in the equidistant separate configuration.}
    \label{fig:five_panel_grid}
\end{figure*}
Thus far, our analysis has assumed an idealized, lossless environment. In realistic implementations, artificial atoms inevitably experience dissipation through spontaneous emission into non-guided environmental modes. To account for this environmental loss, we employ an effective non-Hermitian Hamiltonian approach by making the complex detuning substitution $\Delta \rightarrow \Delta + i\gamma/2$, where $\gamma$ denotes the non-radiative or free-space decay rate of the atom into non-guided modes~\cite{gardiner2004quantum}. For simplicity, we assume identical emitters with uniform environmental coupling ($\gamma_j = \gamma$). Normalizing all parameters by the waveguide decay rate $\Gamma$, the dimensionless loss parameter spans the regime $0 \le \gamma/\Gamma \le 1$. This parameter space restricts our investigation to the physically relevant high-cooperativity regime, where the atom couples more strongly to the guided waveguide modes than to the surrounding environment.

We begin by evaluating the single giant atom configuration under the influence of environmental loss (see Fig.~\ref{fig:five_panel_grid}(a)). Rather than maintaining a fixed detuning, we maximize the routing probability over $\Delta$ to isolate how different spatial scales, parameterized by the phase accumulation $\theta$, are affected by spontaneous emission. As anticipated, dissipation induces a global reduction in routing efficiency. More notably, the point of effective decoupling at $\theta = \pi$ (where transmission to the target waveguide is zero in the lossless limit) broadens significantly into a wide, dissipation-dominated interval as the spontaneous emission rate $\gamma/\Gamma$ increases. As an example of scaling to multi-atom chains, we looked at the fully separated configuration (see Figs.~\ref{fig:five_panel_grid}(b)-(e)). In this setup, the decoupling feature at $\theta = \pi$ remains as a zero-routing band. However, a striking cooperative effect emerges: the width of this unguided decay-dominated interval around $\theta = \pi$ progressively shrinks as the number of giant atoms increases from two to five, demonstrating that collective emitter arrays can structurally suppress the expansion of decoupling zones.

Beyond the behavior at the decoupling node, the multi-atom separate configurations exhibit a remarkable, non-trivial resilience to environmental loss. In the single-atom case, the parameter space allowing high routing efficiency degrades rapidly with increasing $\gamma$. In contrast, the multi-atom chains develop distinct, highly stable horizontal resonance bands (red/orange regions) that maintain near-perfect routing probabilities even deep into the high-dissipation regime ($\gamma/\Gamma \rightarrow 1$). 

In addition to the central decoupling node, the specific phase accumulations $\theta = 0, \pi/2, 3\pi/2,$ and $2\pi$ persist as regions of reduced maximum routing probability. These phase localized dips also exert an adverse proximity effect, dragging down the maximum routing efficiency of neighboring parameters, though this degradation is substantially less severe than the wide-interval suppression observed at $\theta = \pi$. Consequently, the most dissipation-robust operating windows for the equidistant separate layout are restricted to the highly stable phase intervals of $0 < \theta < \pi/2$ and $3\pi/2 < \theta < 2\pi$. Within these specific regimes, the cooperative multi-point interference remains exceptionally resilient against environmental loss across all multi-atom chains.

\begin{figure}
\includegraphics[width=3.4in, height=1.75in]{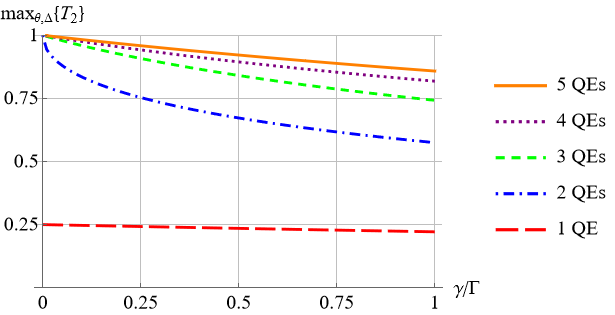} 
\captionsetup{
format=plain,
margin=1em,
justification=raggedright,
singlelinecheck=false
}
\caption{(Color online) Global maximum routing probability $\max_{\theta, \Delta} \{T_2\}$ as a function of the normalized environmental loss rate $\gamma/\Gamma$ for 1 to 5 quantum emitters (QEs). The remaining parameters are the same as those used in the previous plot.}
\label{fig:global_max_routing}
\end{figure}

To explicitly quantify the formation of collective atomic states \cite{albrecht2019subradiant, tiranov2023collective} and cooperative protection offered by scaling the atomic array, Fig.~\ref{fig:global_max_routing} illustrates the global maximum routing probability $\max_{\theta,\Delta}\{T_2\}$ directly against the normalized dissipation rate $\gamma/\Gamma$. While the previous two-dimensional profiles map parameter space landscapes, this global optimization clearly isolates the fundamental limits of the system under worst-case environmental decay. For a single quantum emitter, the routing probability remains strictly capped near $0.25$, demonstrating an inherent vulnerability to symmetry-enforced emission paths. 

In stark contrast, systems featuring multiple giant atoms ($2$–$5$ quantum emitters) achieve perfect routing ($\sim 1.0$) in the lossless limit ($\gamma/\Gamma = 0$). More importantly, increasing the number of emitters systematically decreases the slope of efficiency decay as dissipation grows; with $5$ QEs (solid orange line), the router maintains a remarkable maximum efficiency exceeding $0.85$ even at the upper limit of environmental loss ($\gamma/\Gamma = 1$). This result provides definitive proof that increasing the number of connection points provides a cooperative shield against non-radiative decay, offering a clear advantage for robust quantum-state routing in realistic, lossy waveguide QED architectures without requiring chiral waveguides.

\section{\label{sec:VI} Summary and Conclusions}
In summary, we presented a real-space approach to single-photon routing using an array of up to five giant atoms coupled to two linear waveguides. Each atom was modeled with two spatial connection points per waveguide and uniform coupling strengths. A single giant atom is limited to a maximum routing efficiency of 0.25 due to symmetry-enforced emission paths. Multi-atom arrays (2-5 atoms) introduce rich interference landscapes. Among the configurations examined, the separate and braided topologies enable near-deterministic photon routing (around 1.0), while the highly symmetric nested configuration remains capped at 0.25. Increasing the number of giant atoms expands the operational parameter space for near-perfect transmission. The separate configuration provides the widest and most adaptable parameter ranges for phase accumulation $\theta$ and detuning $\Delta/\Gamma$. The braided configuration achieves comparable optimal performance when the number of atoms is low.

Furthermore, our investigation into atom-atom interactions and non-radiative environmental dissipation highlights the practical viability of these setups. Interatomic interactions break global rotational symmetries, but substantial regions of high-efficiency routing remain in both the separate and braided cases. Under environmental loss ($\gamma < \Gamma$), the separate configuration exhibits remarkable structural robustness, maintaining high-fidelity transmission bands deep into the high-dissipation regime. Although spontaneous emission broadens the sharp decoupling point at $\theta = \pi$ into a zero-routing interval, this degradation is suppressed, and the interval is narrowed as the number of giant atoms increases. Our findings show that collective self-interference among multiple giant atoms enables the engineering of deterministic, tunable single-photon routers in waveguide QED architectures. A promising future direction is to introduce spatial asymmetry by varying the relative distances between giant atomic coupling points or increasing the number of coupling points per giant atom to further optimize multi-port directional routing.

\acknowledgments
The authors acknowledge financial support from the Miami University College of Arts and Science.


\bibliographystyle{ieeetr}
\bibliography{paper}

\end{document}